# Involvement of Surfactant Protein D in Ebola Virus Infection Enhancement via Glycoprotein Interaction

**Anne-Laure Favier [1,2,*], Olivier Reynard [3], Evelyne Gout [4], Martin van Eijk [5], Henk P. Haagsman [5], Erika Crouch [6], Viktor Volchkov [3], Christophe Peyrefitte [1] and Nicole M. Thielens [4]**

[1] Unité de Virologie, Institut de Recherche Biomédicale des Armées, 91220 Brétigny-sur-Orge, France; cpeyrefitte2000@yahoo.fr (C.P)
[2] Unité Imagerie; Institut de Recherche Biomédicale des Armées, 91220 Brétigny-sur-Orge, France
[3] Molecular basis of viral pathogenicity, Centre International de Recherche en Infectiologie (CIRI), INSERM U1111—CNRS UMR5308, Université Lyon 1, ENS de Lyon, 69365 LYON cedex 07, France; olivier.reynard@inserm.fr (O.R); viktor.volchkov@inserm.fr (V.V)
[4] Université de Grenoble Alpes, CNRS, CEA, IBS, F-38000 Grenoble, France; evelyne.gout@ibs.fr (E.G); nicole.thielens@ibs.fr (N.M.T)
[5] Division of Molecular Host Defence, Department of Infectious Diseases and Immunology, Faculty of Veterinary Medicine, Utrecht University, 3512 the Netherlands; M.vanEijk@uu.nl (M.v.E); h.p.haagsman@uu.nl (H.P.H)
[6] Department of Pathology and Immunology, Washington University School of Medicine, St. Louis, MO 63110, USA; crouch@wustl.edu (E.C)
* Correspondence: favier03al@yahoo.fr; Tel.: +33-1-78-65-13-50; Fax: +33-1-78-65-16-46

**Abstract:** Since the largest 2014–2016 Ebola virus disease outbreak in West Africa, understanding of Ebola virus infection has improved, notably the involvement of innate immune mediators. Amongst them, collectins are important players in the antiviral innate immune defense. A screening of Ebola glycoprotein (GP)-collectins interactions revealed the specific interaction of human surfactant protein D (hSP-D), a lectin expressed in lung and liver, two compartments where Ebola was found in vivo. Further analyses have demonstrated an involvement of hSP-D in the enhancement of virus infection in several in vitro models. Similar effects were observed for porcine SP-D (pSP-D). In addition, both hSP-D and pSP-D interacted with Reston virus (RESTV) GP and enhanced pseudoviral infection in pulmonary cells. Thus, our study reveals a novel partner of Ebola GP that may participate to enhance viral spread.

## 1. Introduction

The *Ebolavirus* genus is composed of five species, *Zaire ebolavirus* (type virus, EBOV), *Sudan ebolavirus* (type virus, SUDV), *Tai Forest ebolavirus* (type virus, TAFV), *Bundibugyo ebolavirus* (type virus, BDBV), and *Reston ebolavirus* (type virus, RESTV). EBOV is responsible for severe, often fatal, hemorrhagic fever in humans and nonhuman primates (NHPs) while RESTV is nonpathogenic in humans, but lethal in some NHPs. The last five years have seen the emergence of Ebola outbreaks in unexpected or civil war locations, rendering their control extremely difficult. Since the 2014–2016 EBOV outbreak in West Africa, our perception of the global threat posed by the *Ebolavirus* has changed [1], leading to a better understanding of how EBOV infection takes place [2]. Surprisingly, in numerous cases, when patients fully recovered, the virus was still present in eyes, placenta, semen, breast milk, and lungs, and evidence has emerged that EBOV was able to persist in immune-privileged sites in the body for over several months after its clearance [3–6]. Altogether these findings increased the concerns with regard to control and containment of possible future outbreaks, now including the 2018 outbreak in Congo [7,8]. Moreover, the mechanisms by which the virus causes disease in humans remain insufficiently understood, notably the mechanism leading to tissue invasion by the virus. The role of lectins has been highlighted by several authors and several members of this protein family have been shown to interact with the Ebola glycoprotein (GP) to modify its infectivity. EBOV is known to have a wide cell tropism and cell surface attachment occurs through GP binding to membrane co-receptors, among them lectins (dendritic cell-specific ICAM-grabbing non-integrin/Liver/lymph node-specific intercellular adhesion molecule-3-grabbing integrin, Macrophage galactose binding lectin, Liver and lymph node sinusoidal endothelial cell C-type lectin for DC-SIGN/L-SIGN, MGL, LSECtin, respectively) and other receptors expressed by sensitive cells [9–14]. Previous studies have indicated a role for endogenous circulating mannose-binding lectin (MBL), a member of the collectins family [15], in Ebola infection [16,17]. Depending on the serum conditions, MBL influences Ebola infection, resulting in an enhancement in low complement conditions [18]. In contrast, treatment of mice infected with EBOV using high doses of recombinant MBL had a protective effect [19]. Moreover, independently from the serum complement, a specific interaction involving ficolin-1, a member of the soluble defence collagens family, with EBOV GP resulted in enhancement of virus infection instead of tipping the balance towards its elimination [20].

In this context, we investigated the importance of pulmonary surfactant proteins A (SP-A) and D (SP-D), which play pivotal roles in the innate immune defense of several organs, notably lungs and liver, in EBOV infection. Importantly, SP-A is much more restricted to the lung while SP-D is also present in different mammalian mucosal tissues—including liver, spleen, kidney, lacrimal glands, gastrointestinal tract, and testis [21,22]—organs that are, for some of them, altered during Ebola virus pathology/infection [23].

SP-A and SP-D belong to a family of soluble humoral pattern recognition receptors known as the collectins. These multimeric glycoproteins play an important role in the defense against invading microorganisms, especially in pulmonary tissues. Direct antimicrobial neutralization (binding and aggregation) is often followed by a proinflammatory response to destroy the pathogen [24] and prevent further spread via enhanced phagocytosis of opsonized microbes via macrophages and neutrophils [25,26].

SP-D is a calcium-dependent (C-type) lectin assembled from subunits comprising a C-terminal globular carbohydrate recognition domain (CRD) and a triple helical collagen domain that can multimerize into assemblies of four trimers (dodecamers) and to a lesser extent, depending on pH conditions, into larger oligomers (fuzzy balls) [27]. SP-D is synthesized and constitutively secreted into the airspaces by two types of pulmonary epithelial cells, alveolar type II cells, and Clara cells. SP-D immune activity [28,29] results from its pattern recognition activity towards multiple carbohydrate ligands present on bacteria, fungi, or viruses [30–34]. Differences in the glycan binding specificities of SP-D from different animal species have been reported. Interestingly, specific structural features of the CRD of pSP-D, including a unique sugar binding site and an N-linked oligosaccharide, have been shown to contribute to its distinct activity against influenza A virus (IAV) [35,36].

Many surface viral glycoproteins have been shown to interact with SP-D, notably G and F from respiratory syncytial virus (RSV), HA from IAV, gp120 from HIV, and A27 from vaccinia virus (VACV). In several cases a protective role of SP-D against various viral pathogens has been demonstrated, as for IAV, RSV, and VACV [34,37–44]. At present, there is no evidence for the involvement of those collectins in the innate host defense against EBOV.

The present work characterizes the interplay between EBOV and surfactant defense collectins, more particularly human and porcine SP-D.

## 2. Materials and Methods

*2.1. Cells*

Both Vero E6 cells (Clone E6 of African green monkey kidney cells, ATCC CRL-1586) and human embryonic kidney (HEK) 293T cells (ATCC CRL-1573) were grown in Dulbecco's modified Eagle medium (DMEM) containing 0.11 g/L pyruvate and 4.5 g/L glucose (Gibco). A549 cells (human lung carcinoma cell line, ATCC CCL-185) were grown in F12K medium (Gibco). For the cell culture, media were supplemented with 10% heat-inactivated (56°C, 30 min) fetal calf serum (FCS) (Gibco) and 1% antibiotics (100 U/mL of penicillin and 100 µg/mL of streptomycin, Gibco). Cells were cultured at 37 °C in a 5% $CO_2$ atmosphere.

*2.2. Viruses*

2.2.1. Replicative Viruses

The recombinant vesicular stomatitis virus (VSV) expressing the glycoprotein of EBOV (Mayinga strain) (rVSV-GP) was generated via reverse genetics using a clone of the VSV Indiana serotype containing the GP EBOV open reading frames that were cloned instead of VSV G [45]. rVSV-GP and wild-type EBOV (Mayinga) viruses were produced in Vero E6 cells in DMEM containing 3% FCS. rVSV-GP was propagated under BSL2 conditions and quantified using plaque forming units (PFU). Experiments using Ebola virus were performed in the BSL4 INSERM laboratory Jean Merieux (Lyon, France). EBOV was quantified via plaque assay and revealed using immunohistochemistry (IHC). The day before experimental infection, cells were seeded in multi-wells plates with DMEM medium supplemented with 5% FCS. Before virus infection, cells were rinsed with glucose-free DMEM (Gibco), supplemented with 1% antibiotics. Infection was performed in glucose-free DMEM, in the absence of FCS. For purified EBOV production (Mayinga-EBOV expressing the green fluorescent protein, GFP) [46,47], Vero E6 cells were progressively adapted to grow in a serum-free medium (VPSFM, Life Technologies, Carlsbad, CA, USA) during five passages. EBOV-GFP virus was inoculated at a MOI (multiplicity of infection) of 0.05 and the supernatant was harvested five days post-infection. The supernatant was clarified from cell debris by low speed centrifugation (1500× *g*, 10 min) and then loaded over a 20% sucrose cushion in 10 mM Tris; 150 mM NaCl, 1 mM EDTA, and pH 7.4. Virions were pelleted using ultracentrifugation for 2 h at 134,600× *g* in a SW32 rotor (Beckman Coulter, Fullerton, CA, USA) and the pellet was suspended in 3 mL of phosphate-buffered saline (PBS) containing calcium and magnesium.

2.2.2. Non-Replicative Virus

Recombinant non-replicative vesicular stomatitis virus (VSV) particles expressing the red fluorescent protein (rVSV-RFP) were pseudotyped with RESTV GP (rVSV-RFP-GP-R) as described previously [48].

*2.3. Reagents*

Hemagglutinin (HA) peptide, anti-HA agarose, and rabbit HA epitope tag antibody were purchased from Pierce, Waltham, Massachusetts, USA. Peroxidase-conjugated goat anti-rabbit and anti-mouse IgG antibodies were purchased from Jackson ImmunoResearch, Cambridge, UK. Both mouse monoclonal anti-nucleoprotein (NP) (clone ZDD4) and anti-VP40 (clone 9B2-F2) were produced in-house and diluted at 1:500 and 1:100, respectively. Low viscosity carboxymethylcellulose, mannan, and fatty acid-free bovine serum albumin (BSA) were purchased from Sigma Aldrich, St Louis, MO, USA. True Blue peroxidase substrate was purchased from Seracare KPL (Milford, MA, USA). Protein low binding (LoBind) 1.5 mL tubes were purchased from Eppendorf France SAS (Montesson, France) and 1 M N-2-Hydroxyethylpiperazine-N-2-Ethane Sulfonic Acid (HEPES) solution was from Invitrogen (Carlsbad, CA, USA).

*2.4. Recombinant Proteins*

Recombinant MBL, produced and purified as described previously [49] was kindly provided by NatImmune (Copenhagen, Denmark). Recombinant human SP-D dodecamers were expressed in CHO-K1 cells and purified as previously described [55]. Recombinant trimeric neck + carbohydrate recognition domain fusion proteins (NCRDs) from human (hNCRD and the E321K mutant (mutNCRD)) and rat (rNCRD) species were expressed in bacteria and purified as previously described [50,51]. Recombinant full-length porcine SP-D (pSP-D) was produced in HEK293 cells and purified as described previously [52]. All preparations used for these studies had low endotoxin levels (ranging between 0.27–45.2 ng). AP-SP-A was a kind gift from Dr J.R. Wright (Duke University, Durham, North Carolina, USA). Recombinant human SP-A (rSP-A) was kindly provided by Dr F. McCormack (University of Cincinnati, Cincinnati, Ohio, USA). The molecular size of proteins was estimated as followed: 516 and 600 kDa for hSP-D and pSP-D (composed of twelve identical polypeptides of 43 and 50 kDa, respectively), 72 kDa for both hNCRD and mutNCRD (composed of three identical CRD domains of 24 kDa). Recombinant human ficolin-1 was expressed in S2 insect cells and purified as described previously [53]. Recombinant human ficolin-2 and ficolin-3 were produced in Chinese hamster ovary cells and purified by affinity chromatography on N-acetyl cysteine–Sepharose for ficolin-2 [54] and on acetylated BSA-Sepharose for ficolin-3 [55].

The recombinant GP of EBOV (Mayinga strain) was expressed in 293T cells from pDISPLAY-HA-GP plasmid kindly provided by Pr. E. Ollmann Saphire (Scripps Institute, La Jolla, LA, USA) and purified as described previously [56,57]. Two kinds of trimeric recombinant GPs were used: the transmembrane (TM) domain-deleted protein (residues 33-632; GPΔTM) and the mucin and TM domains-deleted protein (GPΔTM sequence with deletion of residues 312–463, GPΔmucΔTM). The molecular size of soluble monomers was estimated from sodium dodecyl sulfate polyacrylamide gel electrophoresis SDS-PAGE analysis to 150 kDa for GPΔTM [57] and 50 kDa for GPΔmucΔTM [58]. His-tagged recombinant EBOV and RESTV GPs devoid of TM domain and produced in Sf9 insect cells (baculovirus expression system) were purchased from IBT Bioservices, Rockville, Maryland, USA (Z-GPΔTM-b and R-GPΔTM-b). The trimeric nature of the GP recombinant protein was assessed using native PAGE analysis.

*2.5. Interaction of SP-D with GP via an Overlay Assay*

One hundred microliters of purified protein solutions (1 µg/spot) were dotted onto HybondC-extra nitrocellulose membranes (GE Healthcare Lifescience, Pittsburgh, PA, USA). Membranes were blocked for 1 h at room temperature (RT) in Tris-buffered saline containing 0.05% Tween 20, 10 mM CaCl$_2$, and 5% skim milk. The membranes were then incubated overnight at RT in the same buffer containing 2 µg/mL of purified GPΔTM, washed three times for 20 min, and incubated for 1 h at RT with rabbit anti-HA antibody (1/200). After three 20-min washes, the membranes were incubated for 1 h with anti-rabbit horseradish peroxidase conjugate (1/10,000). After three 20-min washes, interaction was detected using a chemiluminescence measurement.

*2.6. Surface Plasmon Resonance Analyses with Immobilized GP Proteins and Data Evaluation*

Surface plasmon resonance (SPR) analyses were performed on a BIAcore 3000 instrument (GE Healthcare, Pittsburgh, PA, USA) at 25 °C. GP proteins and fatty acid-free BSA were diluted to 10 µg/mL in 10 mM sodium acetate, and pH 4.0, and covalently immobilized on CM5 sensor chips in 10 mM HEPES, 150 mM NaCl, 3 mM EDTA, and pH 7.4 containing 0.005% surfactant P20 using amine coupling chemistry, according to the manufacturer's instructions (GE Healthcare). Binding was measured at a flow rate of 20 µL/min in 50 mM triethanolamine-HCl, 145 mM NaCl, and pH 7.4, or in 10 mM HEPES, 150 mM NaCl, and pH 7.4, containing 0.005% surfactant P20 and 5 mM CaCl$_2$. Forty microliters of each soluble analyte at the desired concentrations were injected over the immobilized ligands, and the surfaces were regenerated using 10 µL injections of the running buffer containing 10 mM EDTA and, if needed, 1 M NaCl and 10 mM EDTA. A control flow cell submitted to all coupling steps without immobilized protein or with immobilized fatty acid-free BSA was used as a reference, and the specific binding signal was obtained through subtracting the background signal over the reference surface.

Kinetic data were analyzed via global fitting to a 1:1 Langmuir binding model of both the association and dissociation phases for at least four SP-D concentrations simultaneously, using the BIAevaluation software (version 3.2, GE Healthcare). Buffer blanks were subtracted from the datasets used for kinetic analysis (double referencing). The apparent equilibrium dissociation constants ($K_{DS}$) were calculated from the ratio of the dissociation constant ($k_d$) and association rate constant ($k_a$) ($k_d/k_a$). The values provided were the means ± standard deviations (SDs) from two independent experiments. Although the interaction of oligomeric SP-D with the trimeric GPΔTM was inherently more complex than a simple 1:1 binding model, this model was used for data fitting for comparison purposes and yielded satisfactory chi-square values (<2.5).

*2.7. Virus Infection Assay in the Presence of Collectins*

2.7.1. Vero E6 Experiments

Vero E6 cells were seeded to obtain confluent plates after 24 h of culture in 24-well Multi-Well plates (MW). Replicative viruses (wtVSV, rVSV-GP, and EBOV), and non-replicative GP-pseudotyped particles (rVSV-RFP-GP-R) were incubated with 10 µg/mL of defense collagens (MBL, AP-SP-A, hSP-D, and pSP-D) for 1 h at 37 °C in 10 mM HEPES and 5 mM $CaCl_2$ buffer in LoBind 1.5 mL tubes. During this time, cells were rinsed with fresh glucose-free DMEM containing 1% antibiotics. Then, virus-protein mixtures were incubated with cell monolayers at 37 °C in a 5% $CO_2$ atmosphere during one hour at the following multiplicity of infection (MOI): $5 \times 10^{-4}$ for wtVSV-GP, $1 \times 10^{-4}$ for rVSV-GP, $2 \times 10^{-4}$ for wt-EBOV, $2 \times 10^{-5}$ for purified EBOV-serum free, and 1 for rVSV-RFP-GP-R in 24-well culture plates. For replicative viruses, a low MOI was selected to get a PFU number without loss of resolution after the infection enhancement. Cells were rinsed with glucose-free DMEM and covered with 1.5 mL fresh medium (1:1 carboxymethylcellulose (CMC); DMEM 5% FCS), then cultured for an additional 48 h period and 6 days for VSV and EBOV, respectively. For VSV infection, cells were fixed by adding 0.75 mL of fixing and staining solution (0.2% crystal violet, 4.5% formaldehyde, and 7.5% ethanol in PBS) per well for 2 h. Wells were rinsed twice with water and PFUs were counted. For EBOV infection, the CMC-DMEM mix was removed; cells were fixed by adding a 4% formaldehyde-PBS solution for 10 min and permeabilized by 0.5% Triton X-100 in PBS for 4 min. Then, immunohistochemistry was performed with an anti-nucleoprotein (NP) antibody (1/500) followed by peroxidase-conjugated goat anti-mouse antibody (1/1000). Plaques were visualized by adding 250 µL of True Blue substrate. For the rVSV-RFP-GP-R recombinant virus expressing a reporter gene, cells were covered with fresh medium and cultured for an additional 8-h period and flow cytometry analysis was performed on a Macsquantify VYB flow cytometer (Miltenyi Biotech, Bergisch Gladbach, Germany) as described previously [20]. For each condition, 20,000 events were analyzed, and experiments were performed three times.

2.7.2. A549 Experiments

Similarly, A549 cells were seeded to obtain confluent plates after 24 h of culture in 24-well MW plates. Non-replicative GP-pseudotyped particles rVSV-RFP-GP-R were incubated with collectins. Then, virus-protein mixes were incubated with cell monolayer at 37 °C in 5% $CO_2$ atmosphere during one hour at MOI 1. Cells were analyzed 8 hours post infection on a Macsquantify VYB flow cytometer (Miltenyi Biotech, Bergisch Gladbach, Germany). For each condition, 20,000 events were analyzed, and experiments were performed two times.

*2.9. Statistical Analyses*

For in vitro studies, the two-tailed unpaired Student's *t*-test was performed. Values of $p < 0.05$ were considered significant.

## 3. Results

*3.1. Identification of GP Interaction with SP-D*

The interaction of EBOV GP with human surfactant proteins was first analyzed using an overlay assay. Purified human alveolar proteinosis surfactant protein A (AP-SP-A), recombinant SP-A (rSP-A), and recombinant SP-D (hSP-D) were dotted onto nitrocellulose membranes (with MBL and ficolin-1 as a positive control and ficolin-2 and -3 as the negative controls) and incubated with the soluble trimeric form of GP (GPΔTM). GP binding was detected using a specific anti-HA tag antibody. Apart from ficolin-1 and MBL, previously shown to interact with GP [20], SP-D was the only protein to display a robust binding signal (Figure 1A). AP-SP-A and rSP-A were devoid of GP binding capacity, as observed for ficolin-2 and ficolin-3 (Figure 1A and Reference [20]).

To better characterize those interactions, SPR spectroscopy was used to investigate the interaction of SP proteins with GPΔTM. hSP-D readily bound to immobilized GP in the presence of calcium ions as did the positive control MBL, whereas no interaction was observed for AP-SP-A and rSP-A (Figure 1B), in accordance with the data obtained by the overlay assay. Binding of hSP-D to GP was calcium-dependent, as observed previously for MBL [20], since regeneration of the surface was achieved by injection of EDTA-containing solutions.

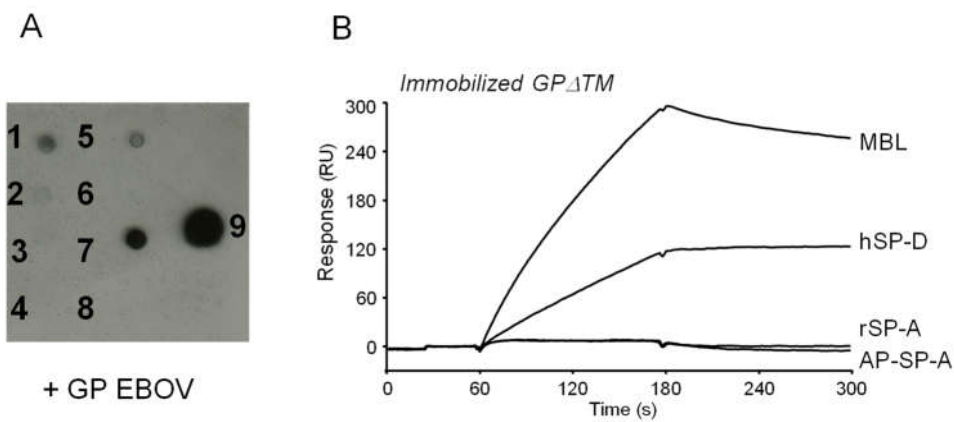

**Figure 1.** Interaction of hSP-D with the GP of EBOV. (**A**) Binding detection via overlay assay. hSP-D was dotted onto nitrocellulose membranes and incubated with 1 μg/mL of purified HA-tagged GPΔTM. After three washes, bound GP was detected with an anti-HA tag antibody and revealed using enhanced chemiluminescence (ECL). GPΔTM (5 ng/spot) and BSA (2 μg/spot) were dotted as positive and negative controls, respectively. 1, hSP-D; 2, AP-SP-A; 3, rSP-A; 4, ficolin-2; 5, ficolin-1; 6, ficolin-3; 7, MBL; 8, BSA; 9, GPΔTM. (**B**) SPR analysis of the interaction of human collectins with immobilized GPΔTM of EBOV. Forty microliters of MBL, hSP-D, rSP-A, and AP-SP-A (2 μg/mL) were injected over 8000 RU of immobilized GPΔTM in 20 mM HEPES, 150 mM NaCl, 5 mM $CaCl_2$, 0.005% surfactant P20, and pH 7.4. The specific binding signals were obtained by subtracting the background signals over a reference surface with 3600 RU of immobilized fatty-free BSA. The results shown are representative of two independent experiments.

*3.2. Characterization of SP-D Binding to GP by SPR Spectroscopy*

SPR spectroscopy was used to further characterize the interaction between hSP-D and GP. Increasing amounts of hSP-D were injected over the immobilized GPΔTM. As displayed in Figure 2A, binding was dose-dependent and further kinetic analysis of the binding data yielded association and dissociation rate constants of $(2.61 \pm 0.92) \times 10^6$ $M^{-1}$ $s^{-1}$ and $(3.19 \pm 0.81) \times 10^{-4}$ $s^{-1}$, respectively, using a global fitting to a 1:1 Langmuir interaction model. The deduced apparent equilibrium dissociation constant ($K_D$) was $0.12 \pm 0.01$ nM, indicative of a high affinity.

EBOV GP is a highly glycosylated protein and the majority of the N-glycosylation sites are concentrated in the glycan cap and mucin-like domain (MLD), while sialylated O-glycans are predominantly located in the MLD. A soluble trimeric recombinant GP, from which MLD was deleted (GPΔmucΔTM), was used to investigate the contribution of the MLD domain in the interaction with SP-D. Dose-dependent binding of hSP-D to the truncated GP was observed and kinetic analysis yielded $k_a$, $k_d$, and $K_D$ values of $(3.65 \pm 0.21) \times 10^6$ M$^{-1}$ s$^{-1}$, $(1.21 \pm 0.08) \times 10^{-3}$ s$^{-1}$, and $0.33 \pm 0.04$ nM, respectively. The SP-D/GPΔTM complex was slightly more stable than the SP-D/GPΔmucΔTM complex, as indicated by a 3–4-fold lower dissociation rate constant and the $k_a$ value was slightly higher in the case of GPΔmucΔTM. Despite these minor differences, these data indicate a high affinity binding of SP-D to both GP forms, suggesting that the mucin domain was dispensable for SP-D binding.

To investigate the role of the carbohydrates in the SP-D/GP interaction, SP-D was injected over immobilized GP in the presence of various carbohydrate ligands. As shown in Figure 2C, SP-D binding was abolished in the presence of 100 μg/mL mannan, and 50% inhibition was observed in the presence of 5 mM mannose or 10 mM N-acetylglucosamine (GlcNAc). The isolated CRD domain also bound to immobilized GP, but the shape of the binding curve was different, with faster association and dissociation rates (Figure 2C). In addition, reduced binding was observed using the NCRD E321K mutant (mutNCRD) (Figure 2D), in which replacement of an acidic residue involved in primary calcium coordination by a lysine residue results in loss of lectin activity [59].

Altogether, these results indicated that the interaction of SP-D with Ebola GP was mediated via the calcium-dependent lectin activity of SP-D towards GP glycans, but not those located in the MLD.

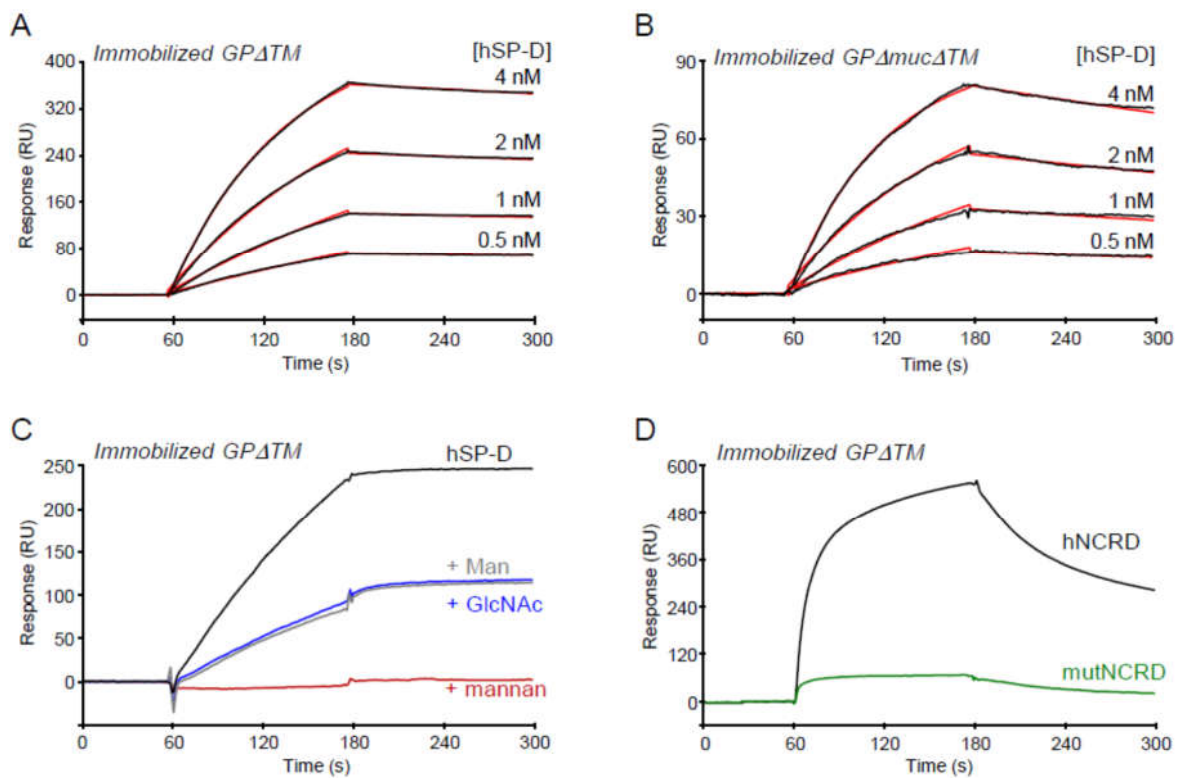

**Figure 2.** Characterization of the interaction of hSP-D with immobilized GPΔTM by SPR. (**A**,**B**) hSP-D samples (40 μL) were injected at the indicated concentrations over immobilized GPΔTM (4,700 RU, panel A) or GPΔmucΔTM (2,500 RU, panel B) in 20 mM HEPES, 150 mM NaCl, 5 mM CaCl$_2$, 0.005% surfactant P20, and pH 7.4 (HBSCa-P). Fits are shown as red lines and were obtained via global fitting of the data using a 1:1 Langmuir binding model. (**C**) hSP-D (3.8 nM) was injected over GPΔTM (8000 RU) in HBSCa-P containing 5 mM mannose (Man), and 10 mM N-acetyl-glucosamine (GlcNAc) or 100 μg/mL mannan. (**D**) hNCRD and its E321K mutant (173 nM) were injected over GPΔTM (8000 RU) in HBSCa-P. (**A**–**D**) The specific binding signals shown were obtained through subtracting the background signal over a reference surface with 3600 RU of immobilized fatty acid-free BSA. The results shown are representative of two independent experiments.

### 3.3. Analysis of hSP-D and pSP-D Binding to Zaire and Reston GP Using SPR Spectroscopy

It has been observed previously that RESTV does not cause disease in humans, whereas its pathogenic potential is known for monkeys and pigs; interestingly in the latter, the lung was identified as a critical replication site [60]. On the other hand, porcine SP-D has been shown to exhibit better hemagglutination activity against influenza A virus than its human counterpart, due to specific glycan binding features in its CRD. To investigate possible differences in the reactivity of both SP-D species with EBOV and RESTV, we compared the binding properties of hSP-D and pSP-D for recombinant GP from Zaïre and Reston Ebola viruses using SPR. Both SP-Ds bound dose-dependently to each GP (Figure 3 A–D) and kinetic analysis of the binding curves yielded apparent dissociation constants ($K_D$) in the nanomolar range (0.26–1.02 nM, see Table 1), reflective of high affinity. However, noticeable differences were observed in the dissociation rate constants, with values of $1.98 \times 10^{-4}$ s$^{-1}$ for hSP-D interaction with EBOV GP and of $8.28 \times 10^{-4}$ s$^{-1}$ for the interaction with RESTV GP, indicating a lower stability of the latter complexes. A similar difference was observed for pSP-D interaction with EBOV and REST. Conversely, the formation of the complexes with RESTV GP was slightly faster than that observed for EBOV GP. The affinity obtained here with recombinant EBOV GP expressed in insect cells and hSP-D (0.26 nM, Table 1) was in the same range as that obtained using the GP expressed in mammalian cells (0.33 nM) (Figure 2A). However, the $k_a$ value was higher for mammalian GP (($3.65 \pm 0.21) \times 10^6$ M$^{-1}$ s$^{-1}$), which might have resulted from the difference in the glycosylation patterns between the recombinant GPs.

Importantly, these in vitro interaction data did not provide evidence for a difference in the binding properties of porcine SP-D for RESTV compared to human SP-D.

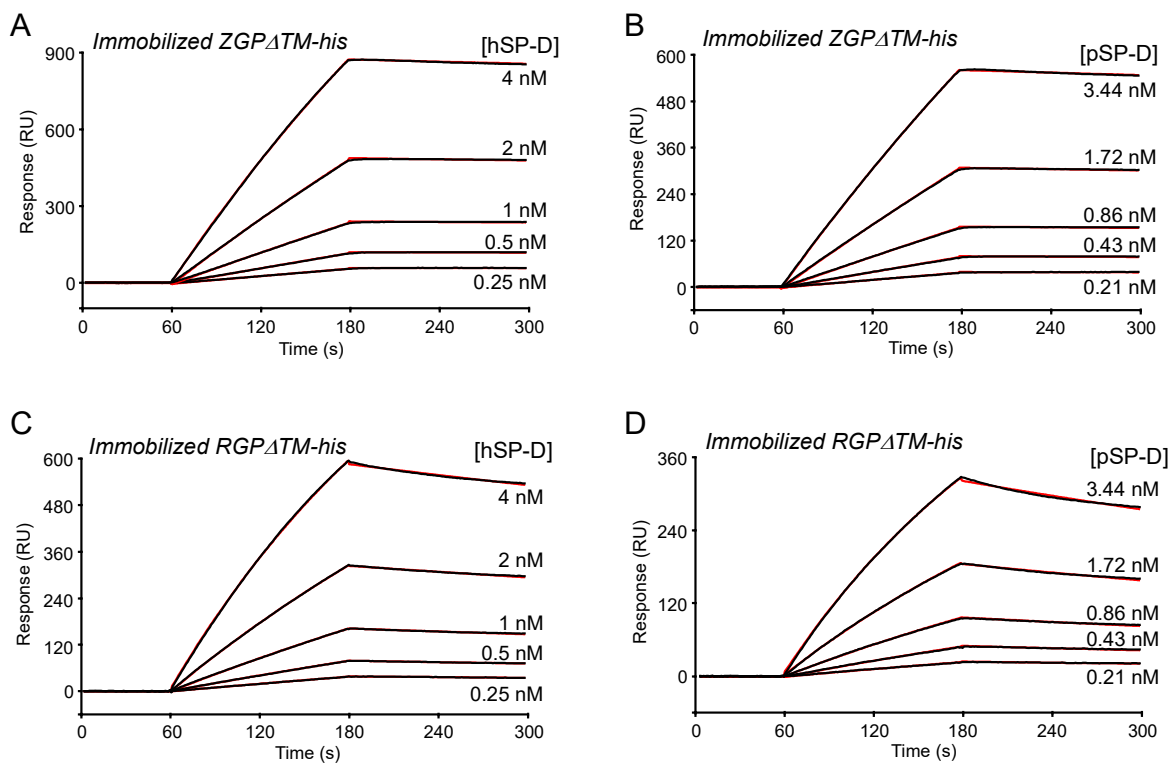

**Figure 3.** SPR analyses of the interaction of human and porcine SP-D with immobilized baculovirus-expressed GPΔTM from Zaïre (ZGPΔTM-b) and Reston (RGPΔTM-b) ebolaviruses. (**A**,**B**) Forty microliters of hSP-D and pSP-D were injected at the indicated concentrations over immobilized ZGPΔTM-b (4800 RU) in 50 mM triethanolamine-HCl, 145 mM NaCl, 2 mM CaCl$_2$, 0.005% surfactant P20, and pH 7.4. (**C**,**D**) hSP-D and pSP-D were injected over immobilized RGPΔTM-b (2,700 RU) in the same buffer. (A–D) Fits shown as red lines were obtained by global fitting of the data using a 1:1 Langmuir binding model. The specific binding signals were obtained by subtracting the background signal over a reference surface obtained through performing the immobilization step without added protein. Each kinetic analysis shown is representative of two independent experiments performed on separate sensor chips.

**Table 1.** Kinetic and dissociation constants for the binding of human and porcine SP-D to immobilized GP from Zaïre and Reston Ebola viruses.

| Immobilized Ligand | Soluble Analyte | $k_a$ (M$^{-1}$ s$^{-1}$) | $k_d$ (s$^{-1}$) | $K_D$ (M) |
|---|---|---|---|---|
| ZGPΔTM-his | human SP-D | $(7.49 \pm 1.10) \times 10^5$ | $(1.98 \pm 0.52) \times 10^{-4}$ | $(2.62 \pm 0.30) \times 10^{-10}$ |
| ZGPΔTM-his | porcine SP-D | $(5.35 \pm 0.42) \times 10^5$ | $(2.28 \pm 0.50) \times 10^{-4}$ | $(4.33 \pm 1.27) \times 10^{-10}$ |
| RGPΔTM-his | human SP-D | $(9.94 \pm 0.37) \times 10^5$ | $(8.28 \pm 0.23) \times 10^{-4}$ | $(8.35 \pm 0.52) \times 10^{-10}$ |
| RGPΔTM-his | porcine SP-D | $(1.29 \pm 0.13) \times 10^6$ | $(1.31 \pm 0.20) \times 10^{-3}$ | $(1.02 \pm 0.08) \times 10^{-9}$ |

Values are expressed as mean ± standard error of the data obtained in two separate experiments on different sensor chips.

*3.4. Both hSP-D and pSP-D Enhance Replicative VSV-GP and EBOV Infection*

The role of the SP-Ds interaction in virus infection was determined using a plaque assay in Vero E6 cells. Initial experiments were performed using a recombinant vesicular stomatitis virus expressing the EBOV GP spike glycoprotein (rVSV-GP) to assay the impact of the collectins on infection. rVSV-GP was preincubated with AP-SP-A, hSP-D, or MBL. Preincubation of the virus with hSP-D resulted in an increase of the virus infection compared to non-preincubated virus ($p = 0.0014$, two-tailed unpaired Student's *t*-test), while no effect was observed with AP-SP-A ($p > 0.05$, two-tailed unpaired Student's *t*-test) (Figure 4A). MBL was used as a positive control of GP interaction ($p = 0.0063$, two-tailed unpaired Student's *t*-test) as described in References [17,20]. hSP-D, preincubated with increasing concentrations (5, 10, and 20 μg/mL), induced a statistically significant dose–response enhancement of rVSV-GP infection compared to the non-preincubated virus ($p = 0.040$, $p < 0.007$ and $p < 0.0001$, respectively) (Figure 4B). pSP-D was assayed in a similar manner, which resulted in an increase of virus infection, as observed for hSP-D ($p < 0.0001$ and $p < 0.0001$, respectively, two-tailed unpaired Student's *t*-test) (Figure 4C). In order to clearly differentiate plaques, a low MOI was used for replicative VSV-GP in these assays, which accounts for a certain variability in the pfu number/well obtained for the basal infection level (in the absence of added collectins).

To confirm this observation, we next investigated whether SP-D mediated enhancement in a real EBOV infection model similarly as observed for rVSV-GP. Incubation of wt-EBOV with increasing concentrations of hSP-D (2.5, 5, and 10 μg/mL) resulted in an increase of virus infection compared with the non-preincubated virus ($p = 0.0002$, $p = 0.0002$, and $p < 0.0001$, respectively, two-tailed unpaired Student's *t*-test), while no effect was observed with AP-SP-A ($p > 0.05$, two-tailed unpaired Student's *t*-test) (Figure 4D). The role of the multimeric nature of SP-D was assayed using a hNCRD construct lacking the multimerization domain. As displayed in Figure 4D, hNCRD lacked the enhancement activity, indicating a critical role of the oligomeric form of SP-D. As the presence of the soluble form of the GP, as well as the presence of serum lectins, may interfere with the collectin binding to GP, the assay was also performed using purified EBOV produced in a serum-free condition. In the same way, preincubation of purified-EBOV with increasing concentrations of AP-SP-A, hSP-D, and pSP-D (1, 5, and 10 μg/mL) resulted in an increase of virus infection for both hSP-D and pSP-D, while no enhancement was obtained for AP-SP-A when compared with the non-preincubated virus ($p < 0.0001$, $p < 0.003$, and $p > 0.05$, respectively, two-tailed unpaired Student's *t*-test) (Figure 4E).

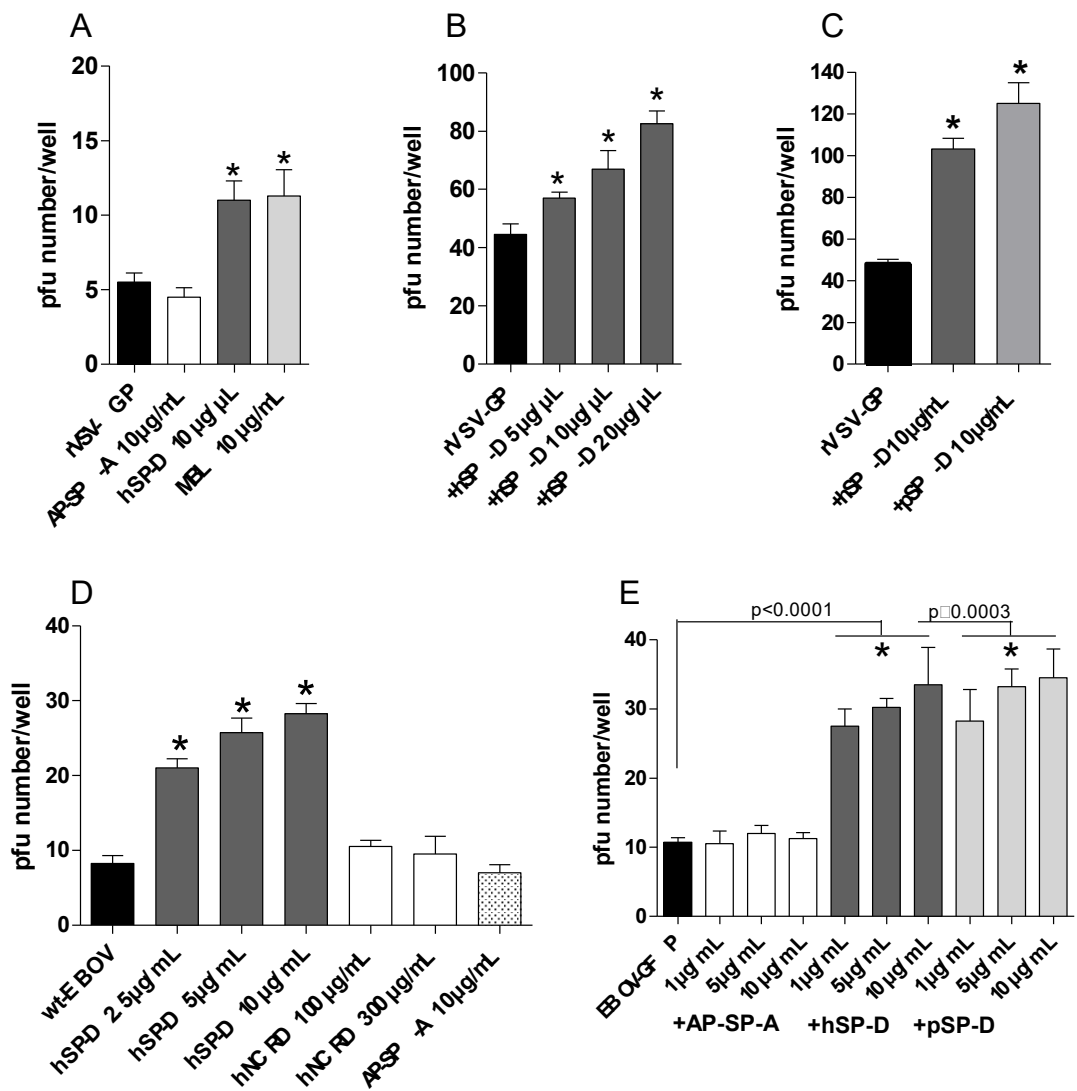

**Figure 4.** Enhancement of replicative-GP virus infection. wtVSV-G, rVSV-GP, and EBOV were preincubated with collectins for 1 h at 37 °C before the infection of Vero E6 cells for 1 h at 37 °C at a MOI of $5 \times 10^{-4}$, $1 \times 10^{-4}$, and $2 \times 10^5$ in 24-well culture plates, respectively. Cells were infected with rVSV-GP preincubated with (**A**) AP-SP-A, hSP-D, and MBL (10 μg/mL), (**B**) increasing concentrations of hSP-D (5, 10, and 20 μg/mL), or (**C**) pSP-D (10 μg/mL). (A, B, and C) At 2 days post-infection, rVSV-GP replication was measured through determination of the PFU number. (**D**) Cells were infected with wt-EBOV preincubated with increasing concentrations of hSP-D (2.5, 5, and 10 μg/mL), hNCRD (100μg/mL and 300μg/mL), and AP-SP-A (10 μg/mL). (**E**) Cells were infected with purified-EBOV preincubated with increasing concentrations of AP-SP-A, hSP-D, and pSP-D (1, 5, and 10 μg/mL). EBOV replication was measured at 6 days post-infection using an IHC assay. The results for the preincubated groups were compared to those for the nonpreincubated groups. *, a statistically difference ($p < 0.05$, two-tailed unpaired Student's *t*-test). The results shown are representative of two independent experiments.

*3.5. SP-D-Mediated RESTV Infection Enhancement in Pulmonary Cells*

Since the routes of infection and the replication site may differ between RESTV and EBOV, we investigated the effect of SP-D on RESTV infection using non-replicative rVSV-RFP-GP expressing the RESTV GP at the surface of the viral particle (GP-R) (Figure 5A).

When SP-Ds were preincubated with rVSV-RFP-GP-R pseudoparticles, only hSP-D enhanced the GP-R pseudotyped particles' infection in VeroE6 ($p = 0.0052$, two-tailed unpaired Student's *t*-test) (Figure 5A) when compared with the non-preincubated particles. Interestingly, pSP-D had no effect with GP-R particles in such a model ($p > 0.05$, two-tailed unpaired Student's *t*-test).

As SP-D is typically synthesized and secreted using pulmonary epithelial cells, we further investigated the capacity of SP-D to enhance pseudovirus infection in a lung epithelium model using the A549 human lung adenocarcinoma cells (Figure 5B). SP-Ds were preincubated with rVSV-RFP-GP-R pseudoparticles, and the percentage of RFP positive cells was analyzed. The capacity of hSP-D to enhance rVSV-RFP-GP-R particles expression was confirmed in this model ($p = 0.0005$, two-tailed unpaired Student's *t*-test). Interestingly, in this pulmonary model cell, and contrary to VeroE6 cells, pSP-D increased the rVSV-RFP-GP-R particles' expression ($p = 0.0002$, two-tailed unpaired Student's *t*-test) compared with the non-preincubated particles. Although similar amounts of viral particles were used to infect VeroE6 and A549 cells, the latter were clearly less infectable than Vero cells, as reflected by a lower percentage of infected cells in the absence of added collectins.

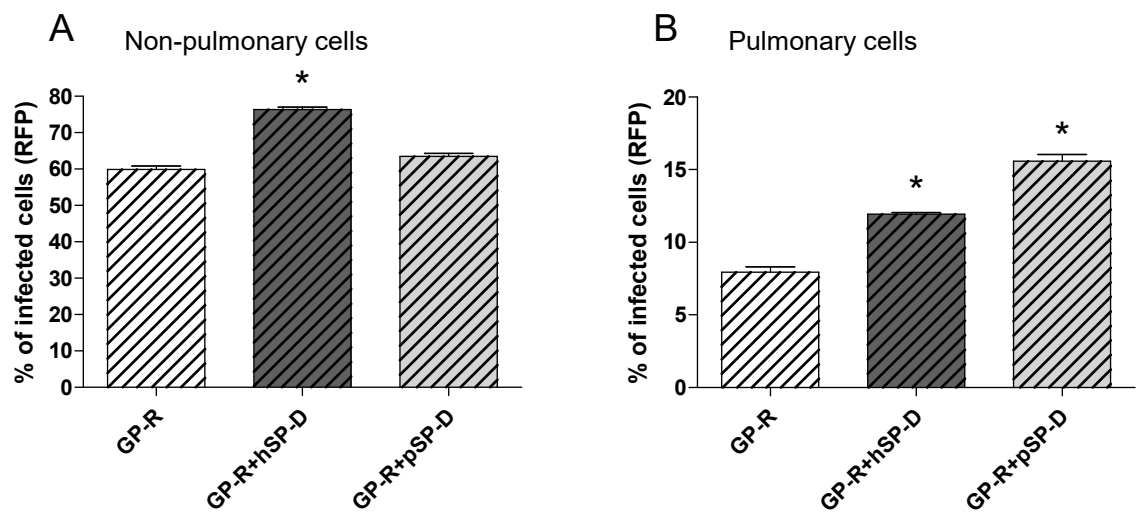

**Figure 5.** Transduction of non-replicative GP-R pseudoparticles in presence of hSP-D and pSP-D. rVSV-RFP-GP-R pseudoparticles were preincubated with hSP-D and pSP-D (10μg/mL) for 1 h at 37 °C before incubation with Vero E6 cells (**A**) or A549 cells (**B**) for 1 h at 37 °C. RFP level expression was analyzed using flow cytometry. The results for the preincubated groups were compared to those of non-preincubated groups. *, a statistically difference ($p < 0.05$, two-tailed unaired Student's *t*-test). The results shown are the mean of three (Vero E6 cells) and two (A549 cells) independent experiments.

## 4. Discussion

The innate immune system plays a critical role in response to viral pathogens and innate immune recognition proteins such as soluble defense collagens, including SP-A and SP-D, are important players in anti-viral defense. Inhibition of IAV using SP-D was the best characterized, but other viruses, such as RSV and VACV, were also reported to be inhibited by SP-D [34,39]. A common inhibition mechanism involves the calcium-dependent interaction of SP-D's CRD domain with sugars localized on the virus spike glycoprotein leading to neutralization of viral infection. However, in rare cases, SP-D was described to facilitate infection, as illustrated for *Aspergillus fumigatus* [61] and *Pneumocystis pneumonia* [62]. In this context, our study shows for the first time that SP-D interaction with a viral glycoprotein can enhance virus infection in mammalian cells.

The interaction between purified EBOV and Reston GP and hSP-D and pSP-D was characterized using SPR analyses. Both hSP-D and pSP-D bound to GP with a high affinity and the interaction involved calcium-dependent binding of the lectin CRD domain to GP glycans. Additionally, the shape of hNCRD-GP SPR binding curve which displays faster association and dissociation rates stresses the importance of the avidity provided by the multimerization of the full-length protein. This avidity seems crucial for infection enhancement as hNCRD, while being able to interact with GP, did not induce infection enhancement. Importantly, we showed that SP-D also interacted with GP exposed at the surface of VSV particles or genuine EBOV, resulting in enhanced infection of Vero cells. While MBL was the first soluble lectin described to bind EBOV GP and to enhance Ebola virus infection in low complement conditions [18], we have recently shown that a second lectin-like protein, ficolin-1, also contributes to the enhancement of EBOV infection, independent of the serum complement level [20]. Ficolin-1 interaction with EBOV-GP was mediated via the fibrinogen-like recognition domain of ficolin-1 and the mucin-like domain of GP through its sialylated moieties residues. In contrast, the interaction of SP-D described in this study was maintained in the absence of the GP mucin domain, as observed previously for MBL [16,18,20]. Interestingly, purified SP-A did not interact with EBOV GP and logically displayed no ability to modulate EBOV infection. The difference between SP-A and SP-D might arise from: i) the different oligomeric organization of SP-A (hexameric, bouquet-like structure) and SP-D (dodecameric, cross-shaped structure), which can result in variations of spatial organization of their trimeric CRDs, influencing the binding for carbohydrate ligand patterns present on EBOV GP; and ii) their differences in sugar binding specificity. The latter hypothesis seems more plausible since MBL, which has an oligomeric organization close to that of SP-A, did interact with EBOV-GP.

Membrane anchored C-type lectins are involved in EBOV infection. Cell surface attachment of EBOV occurs notably through GP binding to membrane lectins (DC-SIGN/L-SIGN, MGL, LSECtin and Myeloid LSECtin) [9–12,14], an interaction that promotes virus entry in various cell types. In this study, SP-D was identified as a new soluble lectin involved in EBOV host cell infection. This result suggests that SP-D likely acts as a multivalent bridging molecule to facilitate attachment of SP-D-bound virus to host cell co-receptors, in accordance with the lack of viral infection enhancement observed when wt-EBOV was preincubated with hNCRD. Candidate collectin receptors on epithelial cells that may interact with the collagen-like regions of SP-D include the calreticulin/CD91 complex [63], the integrin $\alpha 2\beta 2$ [64], and possibly a yet unidentified SP receptor described by Jakel et al. [65]. In addition to facilitating the attachment of the virus particles to host cells, the interaction of virus-bound SP-D to the collagen receptors may have consequences on the modulation of the inflammatory response [66].

Interestingly, hSP-D is also secreted in other parts of the human body as the liver (the major Ebola virus target), spleen, kidney, lacrimal glands, gastrointestinal tract, and testis [21,22]. Most of these sites are known for EBOV replication [23], which raises the possibility that SP-D may also influence infection in several tissues. Interestingly, since SP-D was identified as a new serum biomarker of lung infection [67] or lung injury, it may be useful to assay SP-D serum level as a possible indicator of EBOV pathology progression.

RESTV is unique among ebolaviruses because it does not cause disease in humans [68] or in pig in absence of co-infection [60]. RESTV capacity to infect some animal species exists while the reasons of its non-pathogenicity in humans are not clear. A recent study showed that extended glycans on Reston GP are involved in reduced lectin-mediated viral infectivity of RESTV compared to EBOV [69]. In our study, we did not detect significant differences between hSP-D and pSP-D regarding their capacity to interact with EBOV GP and to enhance infection, which could explain that EBOV replication was observed in the lungs of infected pigs [70–72]. In line with their common capacity to interact with RESTV GP, both hSP-D and pSP-D were able to enhance infection of pulmonary (A549) cells. However, no significant enhancement of infection was observed for pSP-D in the case of non-pulmonary (Vero E6) cells, which was unexpected, given the similar data obtained in SPR experiments for the interaction of pSP-D with the GP of EBOV or RESTV. This suggests that extrapolation of binding data obtained with purified recombinant proteins to a context of cell infection with viral pseudoparticles might be too simplistic. In addition, the different origin of the non-pulmonary and pulmonary cells used (monkey vs human) does not allow direct comparison of the tissue specificity of human and porcine SP-D. Whether our observations reflect differences in the pathogenesis of RESTV in pig remains to be investigated. Future studies should be conducted to increase the knowledge on tissue tropism and the involvement of soluble lectins, notably through the use of ex vivo culture systems of different species and more filoviruses (notably RESTV).

In conclusion, SP-D was identified as a new interacting partner of Ebola GP, contributing in the enhancement of infection instead of providing a first line of defense by inhibiting/neutralizing the virus. Thus ficolin-1, SP-D, and MBL may constitute a viral network of lectin partners used to subvert the innate immune system and promote host cells invasion. Further studies are needed to investigate the underlying mechanisms and the possible role of SP-D in Ebola virus in vivo pathogenesis.

**Abbreviations**

CRD, carbohydrate recognition domain

EBOV, Ebola virus (*Zaïre ebolavirus specie*)

HA, hemagglutinin protein of IAV

IAV, influenza A virus

MBL, mannose-binding lectin

MLV, Murine leukemia virus

NCRD, recombinant trimeric neck+CRDs

SP-A, surfactant protein A

SP-D, surfactant protein D

VLP, virus-like particle

VSV, respiratory syncytial virus

RESTV, Ebola virus (*Reston ebolavirus specie*)

RFP, red fluorescent protein

**Funding:** We acknowledge the support of the Direction Générale pour l'Armement (DGA), the Service de Santé des Armées (PDH-2-NRBC-4-B2-408), and the ARAMI association. This work used the platforms of the Grenoble ERIC-Instruct Center (ISBG; UMS 3518 CNRS-CEA-UGA-EMBL) with support from FRISBI (ANR-10-INSB-05-02) and GRAL (ANR-10-LABX-49-01) within the Grenoble Partnership for Structural Biology (PSB). O.R. and V.V. are supported by the Institut national de la santé et de la recherche médicale (INSERM), Agence Nationale de la Recherche (ANR-14-EBOL-0002-01).

**Acknowledgments:** We thank I. Bally and J.B Reiser for access to the SPR facility and J.P Kleman for access to the cell-imaging platform (flow cytometer). We are grateful to Pr. E. Ollmann Saphire and Dr M. Fusco (Scripps institute, La Jolla, USA) for the kind gift of GP-EBOV Mayinga strain plasmids. We thank Dr. McCormark (Pulmonary and Critical Care Medicine, The University of Cincinnati, Ohio, USA) for the AP-SP-A protein and the late Pr. Jo Wright (deceased on 2012/01/11, Duke University, Durham, North Carolina, USA) for the rSP-A protein.

**Author Contributions:** ALF and N.M.T. conceived and designed the experiments; A.L.F., O.R., and E.G. performed the experiments; A.L.F., O.R., and N.T. analyzed the data; M.V.E., H.H., C.P., and V.V. contributed reagents/materials/analysis tools; A.L.F. and N.M.T. wrote the manuscript draft. All authors approved the final version of the manuscript.

**Conflicts of Interest:** The authors declare no conflict of interest.